\documentstyle[11pt,aasms4]{article}
\begin{document}
\title{
  The Timing Noise of PSR 0823+26, PSR 1706-16, PSR 1749-28, PSR 2021+51 
 and The Anomalous Braking Indices 
 }
\author{Altan Baykal$^{1}$, M.Ali Alpar$^{1}$, Paul E. Boynton$^{2}$,
John E. Deeter$^{2}$,
 }
\affil {
$^{1}$ Physics Department, Middle East Technical University,
Ankara 06531, Turkey, \\ 
$^{2}$ Department of Astronomy, University of Washington, Seattle,
WA, 98195, USA  
 }


\begin{abstract}
We have investigated the stability of 
the pulse frequency second derivatives
($\ddot \nu $) of PSR 0823+26, PSR 1706-16, PSR 1749-28, PSR 2021+51
which 
show significant quadratic trends in their pulse frequency
histories in order to determine whether
the observed second
derivatives are secular or they arise as
part of noise processes. We have used
TOA data extending to more than three decades which are the longest
time spans ever taken into account in pulse timing analyses.
We investigated the stability of pulse frequency second derivative
in the framework of low resolution noise power spectra
(Deeter 1984) estimated from 
the residuals of pulse frequency and TOA data.
We have found that the $\ddot \nu $ terms of these 
sources arise from the red torque noise in the fluctuations of
pulse frequency derivatives which may originate from the 
external torques from the magnetosphere of pulsar.

\end{abstract}
 
\newpage
\section{Introduction}

Rotation powered pulsars emit energy by electromagnetic radiation  
('pulsar braking'). The spin-down is occasionally interrupted by sudden period 
discontinuities ('glitches').    
In addition, pulse arrival time measurements display
 irregularities in the rotation rate 
known as 'timing noise'.  
The timing noise above the measurement errors 
could be due to a noisy component of the secular torque 
involving fluctuations in the magnetosphere  
of the neutron star (Cheng 1987 a,b; 1989).
Alternatively, timing noise could arise   
from internal torques coupling different components of the 
neutron star, for example the de/recoupling of the crust superfluid 
(Alpar, Nandkumar and Pines 1986, 
Jones 1990). 
Timing noise for pulsars were studied in last three decades  
(Boynton et al., 1972, Groth 1975, 
Cordes 1980, Cordes and Helfand 1980,
 Cordes and Downs 1985, D'Alessandro et al., 1995, 1997, 
Deshpande, et al., 1996). 
Boynton et al. (1972) proposed that the timing noise in the 
times of arrival (TOA) of pulses    
might arise from 'random walk' process which are 
r$^{th}$ order (r=1,2,3)
time integrals  
of a 'white noise' time series 
 (or time series of unresolved delta functions). 
The random walks in phase $\phi $, pulse frequency $\nu$ and 
pulse frequency derivative $\dot \nu$ are called   
'phase noise', 
'frequency noise' and 'slowing down noise' respectively 
(Cordes 1980).    
The timing noise of the Crab pulsar was found to be consistent with 
'frequency noise' 
(or random walk in pulse frequency) (Boynton et al., 1972, Groth 1975). 
The timing 
analyses of other pulsars' time series   
showed that they are also  
consistent with random walk processes which are superposed on
identifiable micro-jumps in the timing parameters
(Cordes 1980 and Cordes and Downs 1985).

The cross talk between the 
timing noise and secular slowing down is very important.  
 Many of the middle aged pulsars with spin-down age 
$\tau = P/2\dot P$ greater than about 
10$^{6}$ years have shown 
anomalous trends in their secular frequency second derivative ($\ddot \nu$)
(Cordes and Downs 1985). 
 These 
trends make it impossible to recover the braking law 
$\dot \nu \sim  \nu ^{n} $ of the pulsar 
(for pure magnetic dipole radiation n=3).
 Nominal values of $\ddot \nu$ from timing fits gave anomalous   
 braking indices ranging from $-10^{4}$ to $10^{4}$ in various pulsars.
 Recent observations 
of some young/middle aged pulsars with glitches also    
showed anomalous     
positive braking indices of the order $\sim $20 -200 (Shemar and Lyne 1996). 
 Interglitch recovery between successive glitches can effect 
the pulsar's dynamical parameters such as $\dot \nu$ and $\ddot \nu $.
For the glitching pulsars, the high values of the second derivative
of the rotation rate, $\ddot \nu $, 
and associated braking indices of order 20-200 
are characteristic of interglitch 
recovery 
 (Alpar 1998), 
which extends from one glitch to the next one, 
as studied in detail between the glitches of the Vela 
pulsar (Alpar et al., 1993).
For all middle aged pulsars the expected intervals between glitch events are of the 
order of a few hundred years  
 (Alpar and Baykal 1994). 
Thus a pulsar is most likely to be observed during the 
interglitch recovery phase.   
This raises the question of whether other middle aged 
pulsars also exhibit typical interglitch recovery or  
whether timing noise dominates the observed timing of these pulsars.

We have investigated the time series of 
 pulsars on the longest available time scales by combining the earlier 
observations of 24 pulsars (Downs and Reichley 1983) and 
recent observations (Siegman, Manchester and Durdin 1993, 
Arzoumanian, Nice and Taylor 1994). 
In this way, 
 we have available timing data for time spans of the order of 30 years 
for several pulsars.  
Several of these pulsars are automatically eliminated 
as candidates for secular timing behaviour, since their 
frequency time series are not consistent with secular quadratic trends 
(constant $\ddot \nu$). Alternatively, polynomial fits to the TOA of 
these pulsars can  
require higher order polynomials rather than a cubic polynomial.
For these pulsars the time series is dominated by complicated 
noise processes rather than interglitch recovery. 
For four pulsars, PSR 0823+26, PSR 1706-16, PSR 1749-28, and 
 PSR 2021+51, 
there are significant quadratic trends in frequency histories 
(cubics in TOA). 
In this work, we evaluate these quadratic trends 
to decide whether they  
are secular trends  
 or  
 just part of the timing noise process. 

In Sec 2., we describe the TOA
 of the pulses and present 
pulse frequency histories of the four pulsars. 
In Sec. 3, we construct the power density spectra in the fluctuations 
of derivatives of pulse frequencies using the mean-squared 
residuals technique developed by Deeter (1984) and compare the slowing 
down parameters of the pulsars with the same parameters deduced 
from noise strengths.

\section {The data bases} 

The TOA of pulses
used in this work are drawn from the observations of 
Downs and Reichley (1983) 
from
1968 to 1982
using the Goldstone telescope in California;   
Siegman, Manchester and Durdin (1993) 
from 1986 February to 1988 July, using the
Molonglo Observatory Synthesis Telescope 
and Arzoumanian, Nice and Taylor (1994),
from
1989 August to 1990 February, using the
National Radio Astronomy Observatory
in Green Bank West Virgina. 
The TOA of pulses which were transformed by Downs and Reichley (1983) 
and Siegman, Manchester and Durdin 1993 to geocenter and 
Arzoumanian, Nice and Taylor (1994) to topocenter, were 
transformed to the solar 
system barycenter.

 Pulse cycle counts were assigned to the 
recorded TOA of pulses. In this procedure data sets 
separated by gaps which are 
of the order of several years were   
not combined into single ephemerides, 
to avoid possible cycle count 
ambiguities (note that we estimated the power spectra 
in low frequency  
by using the pulse frequency time series).  
The local pulse frequencies are estimated by fitting 
the cycle counts with a Taylor series        
\begin{equation} 
\phi = \phi_{0} + \nu _{0} (t-t_{0}) + \frac{1}{2} \dot \nu (t-t_{0})^{2}
\end{equation} 
where $t_{0}$, 
$\phi_{0}$, $\nu _{0}$, and $\dot \nu $ are the  
time ephemeris, phase, frequency and derivative of the frequency. 
Typically, these fits are determined on data sets of 
$\sim 10^{2}$ days.
This timescale is relatively 
short for the apperance of red noise components and possible 
braking indices
in the residuals of TOA of pulses 
($\ddot \nu /6 (t-t_{0}) ^{3} << 1 $). 
We therefore   
did not include the cubic terms in our fits. 
Table 1 presents the timing solution of 
PSR 0823+26, PSR 1706-16, PSR 1749-28, PSR 2021+51, from the 
pulse frequency history records.  
In Fig. 1, the long term history of pulse frequency residuals 
is represented by removing 
the long term spin-down trend $(\dot \nu )$. 

\section{ Power Spectra} 

 The technique applied in this work for the estimation 
 of red noise power density and associated random walk noise 
 strengths is discussed in detail by Deeter and Boynton (1982) and 
 Deeter (1984). Some aspects are also 
 summarized here. In the case of rth-order red noise with 
 strength S$_{r}$, the mean square residual for data spanning an interval 
 with length T is proportional to S$_{r}$T$^{2r-1}$. 
 The proportionality factor depends on the degree m of the polynomial 
 removed prior to computing the mean square residual; this factor 
 can be obtained by determining the expected mean square residual 
 for unit strength red noise S($_{r=1}$) over unit interval (T=1), 
 either by Monte-Carlo methods (Cordes 1980) or by direct mathematical 
 evaluation (Deeter 1984). The expected mean square residual, after 
 removing a polynomial of degree m over an interval of length T, 
 is given by 

\begin{equation} 
<\sigma _{R}^{2}(m,T)> = S_{r}T^{2r-1}<\sigma _{R}^{2}(m,1)>_{u},
\end{equation} 
where the subscript u indicates that the expectation 
has been derived for a unit-strength noise process. 
We estimated the noise strengths at the longest time 
scale T$_{max}$ or at lowest frequency; f=1/T$_{max}$, where 
 T$_{max}$ is the maximum time span of the data, from the residuals 
of pulse frequencies by removing their linear trends.
For the shorter intervals or higher frequencies; 
f$_{n}$=n/T$_{max}$, where n is a positive integer, 
we removed quadratic 
trends from the TOA data. 
In order to see whether the noise strengths are     
stable or not and to see whether the 
quadratic trends in pulse frequency 
and  
cubic trends in TOA absorb the noise, we estimate alternative sets of 
noise strengths by removing quadratic polynomials 
from the pulse frequency data for the longest time span of data 
and cubic polynomials  
from the TOA data for the shorter intervals.
 In Figure 2, noise strengths (or power 
spectra estimates) are plotted against the reciprocal of the time scale 
(or sampling frequency 
 f=n/T). It is found that 
for each source these two 
power spectra are consistent with each other 
in terms of average noise strength 
$S_{r}$ and slope of the power spectra.
This shows that our original noise estimates 
 were robust, 
(consistent with each other in terms of the noise strength parameter, 
$S_{r}$) and  
were not dominated by either of the two particular polynomial trends. 
If there were a secular polynomial trend in the data, we could 
expect that particular trend to produce a significantly better fit, i.e. 
a significantly lower, and different power spectrum, compared to the 
other polynomial models.   

\subsection{Diagnostic of Power Spectra }
In the estimation of noise strengths, we adopted 
r=1 for pulse frequency data and r=2 for the TOA data.
If the noise strength measurements are 
constant, independent of observing frequency or
equivalently, the power specra have zero slope, then 
pulse frequency fluctuations can be explained in terms of 
a stationary random walk model and fluctuations of the pulse frequency 
derivatives can be 
explained as a stationary white noise model: 
at all timescales the same noise process prevails. For the data 
from the four pulsars at hand, the power 
density estimates (or noise strengths) have higher values at low frequency.
In the power spectra, 
power density estimates in the derivative 
of pulse frequencies (see Fig. 2, and Tab. 2) 
behave approximately as P$_{\dot \nu} \sim f^{- \alpha }$ 
where $\alpha \sim$ 0.5-2.4, indicating the presence of strong red 
noise. As seen from Table 2 
two of the sources PSR 0823+26 and PSR 1749-28 have   
non-even integer power-law indices with large error bars 
in the power spectra of their pulse frequency derivatives. 
On the other hand, PSR 2021+51 and PSR 1706-16 have power-law indices 
$\sim $ -2 which implies that pure random walk in 
the pulse frequency derivatives (or P$_{\dot \nu } \sim f^{-2}$) 
(Boynton and Deeter 1982) is possible.
This suggests either some of these pulsars experience unresolved  
step like perturbations which are superposed on white noise process   
in their pulse frequency derivative 
histories (mixed process) or they are making pure random walks in their 
pulse frequency derivatives.
The deviation from white noise to red noise in 
pulse frequency derivative may arise from both arguments. 
  
\section{Conclusion}
We have   
investigated the stability of 
the pulse frequency second derivatives      
($\ddot \nu $) of four pulsars 
which  
show significant quadratic trends in their pulse frequency 
histories in order to determine whether 
the observed second 
derivatives are secular or they arise as 
part of noise processes. We have used  
TOA data extending to more than three decades. These are the longest 
time spans ever taken into account in pulse timing analyses.
We investigated the stability of pulse frequency second derivative  
in the framework of low resolution noise power spectra 
(Deeter 1984) estimated 
from  
the residuals of pulse frequency and TOA data. 
Two independent noise strengths were estimated for each source,   
by removing the $\dot \nu$ and $\ddot \nu $ from the longest time span 
of data sets and 
their independent subsets of data.  
We found that the power spectra 
obtained from the residuals after removing 
$\dot \nu$ or an additional secular $\ddot \nu $ term were 
 consistent 
each other at the $1\sigma$ confidence level. This shows that the 
quadratic trends ($\ddot \nu $) in pulse 
frequency time series were not secular:
in the sense that including a $\ddot \nu $ term in the 
fits does not provide a significant improvement over the fits 
with $\dot \nu$ alone. 
 
What are the possible 
sources of the   
noise process on these systems?    
Using the vortex creep theory, Alpar et al., (1986) developed 
model noise power spectra for three different types of events 
which might cause to observed timing noise. 
These events were: (i) pure vortex unpinning events 
(micro glitches), (ii) breaking of the crustal lattice by pinned 
vortices, (iii) external events which do not involve vortex unpinning.  
They predict the form of the power 
spectrum of timing noise in the derivative of 
pulse frequency ($\dot \nu $) to be 
\begin{equation}
P_{\dot \nu}(f)=4\pi^{2}
 R<\Delta \Omega ^{2}> \frac{(1-Q)^{2}+f^{2}\tau^{2}} 
{4\pi^{2}+f^{2}\tau^{2}},~rad^{2}~sec^{-3} 
\end{equation}     
where R is the micro glitch or breaking of crust rate (event rate), 
$<\Delta \Omega ^{2}>$ root mean square of 
the pulse frequency fluctuations (step size of the fluctuations),        
$Q=\frac{\int_{0}^{\inf}
\Delta \dot \nu (t) dt}{\Delta \nu}$ is the fractional change 
in the pulse frequency for each relaxed event, $\tau$ is the 
relaxation time for each event.

 The above model 
noise spectra change at around  
f=$\tau ^{-1}$. Q=1 is the 
pure unpinning case. In this model, logarithmic slope 
of the spectrum changes from +2 (blue) to 0 (white) noise 
at f=$\tau ^{-1}$. If $(1-Q)^{2} >>$1 or $(1-Q)^{2}<<$1 
the events are dominated by crust breaking. 
Initial pulse frequency jump 
can relax to any value due to the initial 
crust quake. Therefore Q can take any value
different from 1. The power spectrum of the noise resulting 
unpinning events from initial crust quake (or mixed events) 
have slopes 0 at low frequency, $\pm$2 at $f \sim \tau ^{-1}$, 
and 0 at high frequency. Therefore white noise in the  
$\dot \nu$ time series is expected at very low and high frequencies. Change 
in the noise level is expected at time scales $\tau$ 
due to the coupling of superfluid layers. 
Naturally events which do not involve 
vortex unpinning, should show almost null coupling 
(or inifinite coupling time), 
therefore flat noise spectra  
in the fluctuations of $\dot \nu$ time series is expected. 
However, the power spectra at hand favor red noise in 
$\dot \nu$ fluctuations at all frequencies especially 
at low frequency.

 Cheng (1987a,b) has suggested that red noise 
in pulse frequency derivatives may arise from external 
torques from the magnetosphere of pulsar. In this model, pair 
production processes in the outer gap of magnetosphere can give 
rapid variation in the current braking torque 
$\delta N_{\delta J\times B}$ where $\delta J$ is a perturbed
 current on the 
magnetosphere of pulsar and $B$ is the external magnetic field. 
If the current braking torque perturbs  
the rotation rate of the neutron star 
by microglitches 
this torque 
will remain unchanged until the next microglitch and 
hence, step-like changes in the torque will give rise 
to a random walk in $\dot \nu$. 
The 
rate of the torque variations, naturally, 
will be the same as the rate of microglitches 
which may occur due to the 
internal re/de coupling of crust superfluid and crust 
or sudden movement of neutron star crustal plates 
of the type described by Ruderman (1991).  
Of course if the micro glitches occur seldomly, 
a quasi-random walk in the power spectra 
of $\dot \nu$ time series is expected. Our measured noise power spectra 
in the $\dot \nu$ time series are consistent with red noise 
and hence may arise from magnetospheric torque 
fluctuations. 

{\bf Acknowledgements~:}\\
This work is supported by Nato Collaborative Research Grant (931376) 
and The Scientific and Technical Research Council of Turkey, 
T{\"U}B{\'I}TAK, under Grant TBAG {\"U}-18. MAA acknowledges 
support from the Academy of Sciences of Turkey. 
  
\hspace{-1.1cm}{\Large{\bf References}}

\hspace{-1.1cm} Alpar, M.A., 1998, Adv. Space. Res., {\bf 21}  
                159  

\hspace{-1.1cm} Alpar, M.A., Baykal, A., 1994, MNRAS, {\bf 269,} 
                849 

\hspace{-1.1cm} Alpar, M.A., Chau, H.F., Cheng, K.S., 
                Pines, D., 1993, ApJ., {\bf 409,} 345 

\hspace{-1.1cm} Alpar, M.A., Nandkumar, R., Pines, D., 
                1986, ApJ., {\bf 311,} 197

\hspace{-1.1cm} Arzoumanian, Z., Nice, D.J., Taylor, J.H., 
                1994,{\bf 422,} 671 

\hspace{-1.1cm} Boynton, P.E., Groth, E.J., Hutchingson, D.P.,
                Nanos, G.P., Partridge, R.B., Wilkinson, D.T.,
                1972, ApJ., {\bf 175,} 217

\hspace{-1.1cm} Cheng, K.S. 1987a, ApJ., {\bf 321,} 799 

\hspace{-1.1cm} Cheng, K.S. 1987b, ApJ., {\bf 321,} 805 

\hspace{-1.1cm} Cheng, K.S. 1989, in Timing Neutron Stars, Eds. H.{\"O}gelman, 
                E.P.J. van den Heuvel, pp. 503-509   
               
\hspace{-1.1cm} Cordes, J.M., 1980, ApJ., {\bf 237,} 216 

\hspace{-1.1cm} Cordes, J.M., Downs, G.S., 1985, ApJ. 
                Suppl. Ser., {\bf 59,} 343  

\hspace{-1.1cm} Cordes, J.M., Helfand, D.J., 1980, 
                ApJ., {\bf 239,} 640

\hspace{-1.1cm} Deeter, J.E., Boynton, P. E., 
                1982, ApJ., {\bf 261,} 337  

\hspace{-1.1cm} Deeter, J.E., 1984, ApJ., {\bf 281,} 482  

\hspace{-1.1cm} Downs, G.S., Reichley, P.E., 1983, 
                Astrophys. J. Suppl. Ser., 
                {\bf 53,} 169 

\hspace{-1.1cm} D'Alessandro, F., McCulloch, P.M., Hamilton., P.A., 
                Deshpande, A.A., 1995, MNRAS, {\bf 277,} 1033 

\hspace{-1.1cm} D'Alessandro, F., Deshpande, A.A., McCulloch, P.M., 
                1997, J.Astrophys. Astr. {\bf 18,} 5 

\hspace{-1.1cm} Deshpande, A.A., D'Alessandro, F., McCulloch, P.M., 
                1996, J.Astrophys. Astr. {\bf 17,} 7 

\hspace{-1.1cm} Groth, E.J., 1975, Astrophys. J. Suppl. Ser., 
                {\bf 29,} 443

\hspace{-1.1cm} Jones, P.B., 1990, MNRAS, {\bf 246,} 364 

\hspace{-1.1cm} Ruderman, M. 1991, ApJ., {\bf 382,} 587                

\hspace{-1.1cm} Shemar, A.L., Lyne, A.G., 1996, MNRAS, {\bf 282,} 677 
 
\hspace{-1.1cm} Siegman, B.C., Manchester, R.N., Durdin, J.M., 
                1993, MNRAS, {\bf 262,} 449 

\newpage
\begin{table}
\caption{Timing Solution of PSR 0823+26, PSR 1706-16, PSR 1749-28, PSR 2021+51, from pulse frequency history 
   }
\begin{center}
\begin{tabular}{| c| c|}  \hline
PSR 0823+26 & \\  \hline  
Epoch(JD)  & 2 443 754.5472134(6)       \\ \hline
Pulse Frequency (Hz) & 1.88444538582(4)  \\ \hline
Pulse Frequency Derivative (Hz s$^{-1}$) & -0.60494(2)$\times 10^{-14}$
  \\ \hline 
Second Derivative of Pulse Frequency 
(Hz s$^{-2}$) & -0.208(3) $\times 10^{-24}$ \\ \hline \hline 
PSR PSR 1706-16 & \\ \hline  
Epoch(JD)  & 2 444 455.1161243(5)        \\ \hline
Pulse Frequency (Hz) & 1.53127025825(9)  \\ \hline
Pulse Frequency Derivative (Hz s$^{-1}$) & -0.14888(2)$\times 10^{-13}$
 \\ \hline  
Second Derivative of Pulse Frequency
(Hz s$^{-2}$) &  0.2517(7)$\times 10^{-24}$ \\ \hline \hline
PSR 1749-28 & \\ \hline  
Epoch(JD)  & 2 442 622.66650772(9)        \\ \hline
Pulse Frequency (Hz) & 1.77760420787(4)  \\ \hline
Pulse Frequency Derivative (Hz s$^{-1}$) & -0.25714(5)$\times 10^{-13}$
 \\ \hline  
Second Derivative of Pulse Frequency
(Hz s$^{-2}$) &  0.112(1)$\times 10^{-24}$ \\ \hline \hline
PSR 2021+51 & \\ \hline  
Epoch(JD)  & 2 443 488.7150164(5)       \\ \hline
Pulse Frequency (Hz) &  1.88965873165(7)   \\ \hline
Pulse Frequency Derivative (Hz s$^{-1}$) &
 -0.109195(3)$\times 10^{-13}$ \\ \hline
Second Derivative of Pulse Frequency
(Hz s$^{-2}$) &  -0.56(3)$\times 10^{-25}$ \\ \hline \hline

\end{tabular}
\end{center}
() denotes 1 sigma error according to least square 
fitting (red noise contribution in the error estimates are not included) 

\end{table}

\begin{table}
\caption{Parameters of Power Spectra} 
\begin{center}
\begin{tabular}{| r| c| c |}  \hline
Pulsar      &Mean Noise Strength         & power law               \\ \hline
            & log( sec$^{-3}$)  & index                    \\ \hline   
PSR 0823+26(a)&  -25.93                  & -1.16 $\pm$ 0.48          \\ 
           (b)&  -25.93                  & -0.39 $\pm$ 0.36       \\ \hline 
PSR 2021+51(a)&  -26.21                  & -1.95 $\pm$ 0.74      \\
           (b)&  -26.72                  & -1.64 $\pm$ 1.40      \\ \hline 
PSR 1706-16(a)&  -25.65                  & -2.41 $\pm$ 0.79      \\
           (b)&  -25.88                  & -1.94 $\pm$ 1.33      \\ \hline 
PSR 1749-28(a)&  -26.03                  & -0.88 $\pm$ 0.50      \\ 
           (b)&  -26.27                  & -1.05 $\pm$ 0.83      \\ \hline 

\end{tabular}
\end{center}

(a) estimates the noise strengths from the  
 residuals from linear trend of pulse frequency data and 
 quadratic trend of TOA data. (b) uses one order higher 
 degree of polynomial in the estimation of 
 noise strengths (see the text). 

\end{table}

\newpage 

{\Large{\bf Figure Caption}}\\

{\bf Fig. 1}~Pulse frequency residuals of PSR 0823+26, 
PSR 1706-16, PSR 1749-28, PSR 2021+51 after removing secular spin-dow 
trend of the data \\

{\bf Fig. 2}~Power density spectra of the fluctuations in the pulse 
frequency derivatives ($P_{\dot \nu}$) of PSR 0823+26(a,b), 
PSR 1706-16(a,b), PSR 1749-28(a,b), 
PSR 2021+51(a,b). The power 
density estimates are indicated by error bars. Vertical error bars 
indicate the stability of each power estimate (approximately 
$\pm 1\sigma$ error bars) as obtained by 
Monte-Carlo simulations inculding the 
uncertainties in the pulse frequency and TOA 
data. 
Horizontal error bars indicate approximate $\pm 1\sigma$ error bars 
of the distribution of analysis frequencies sampled by each power 
density estimate, as specified by Deeter (1984). The contributions 
to the noise strengths due to uncertainties in the pulse frequency and TOA 
data are plotted crosses. The notation (a) denotes that in power 
density estimates are obtained by removing, linear trends
from pulse frequency data and quadratic trends 
from the TOA data and (b) denotes
 that one degree higher polynomial are removed each   
power density estimate.\\    


\newpage
\clearpage
\begin{figure}
\plotone{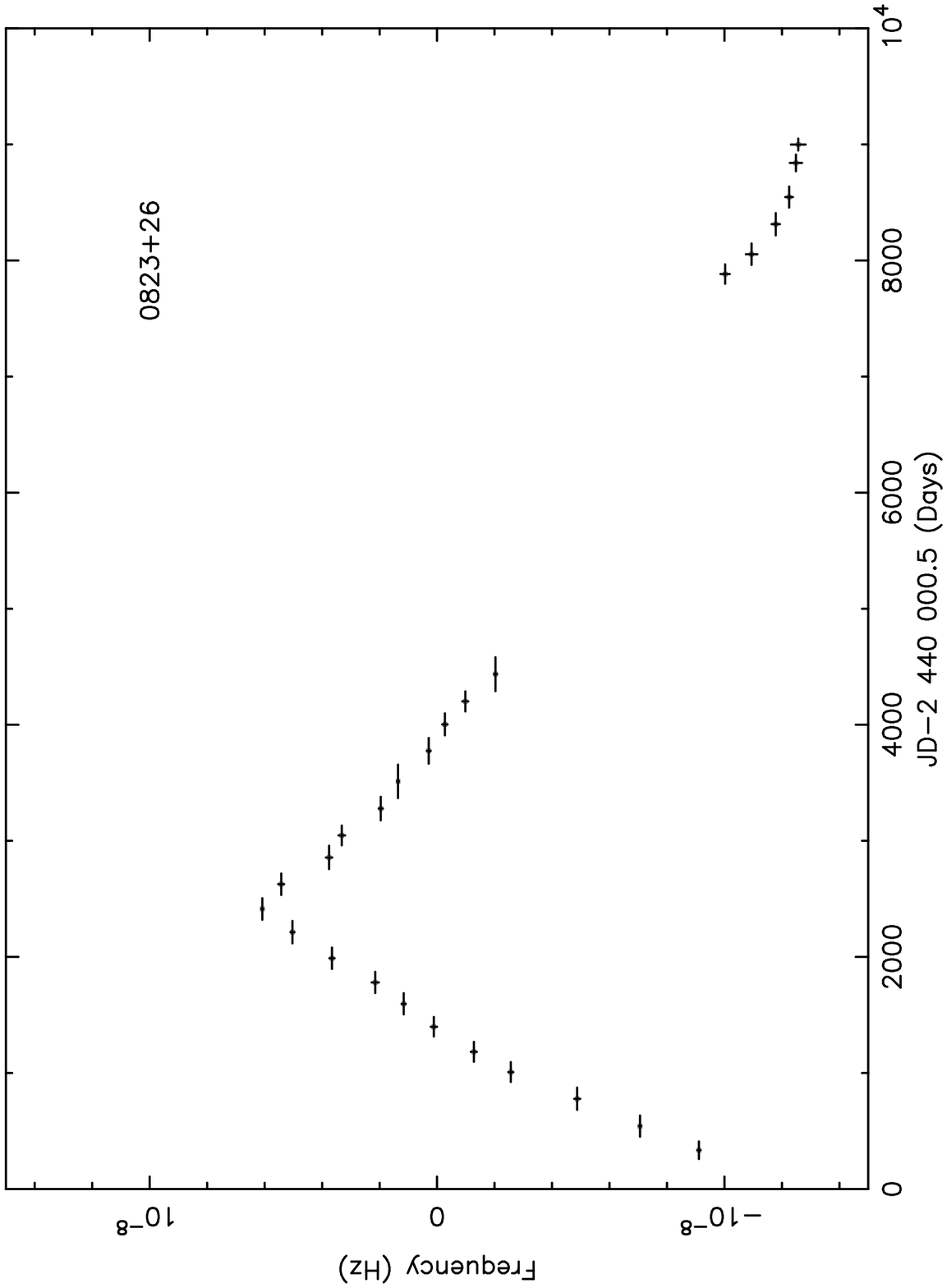}
\end{figure}
\newpage
\clearpage
\begin{figure}
\plotone{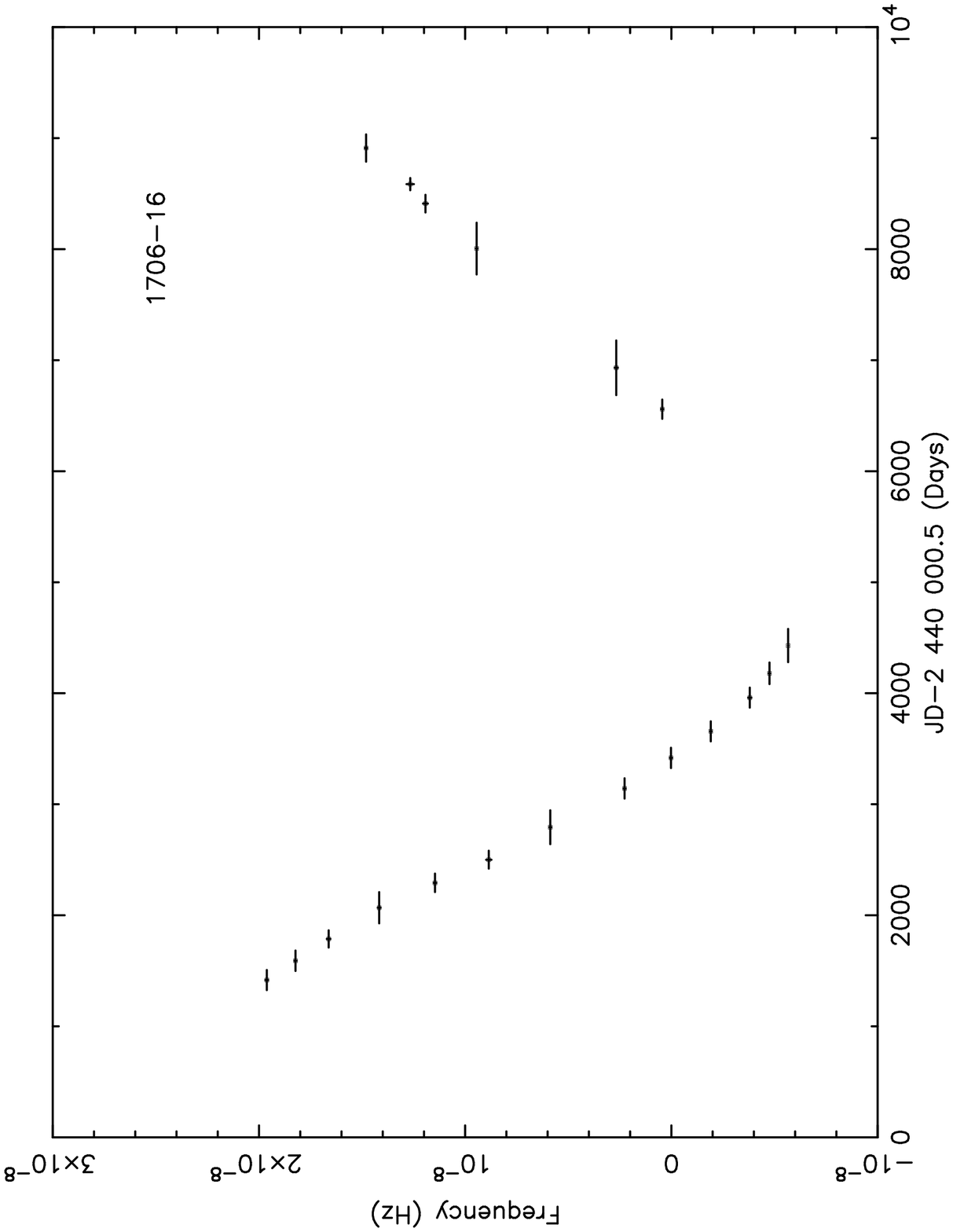}
\end{figure}
\newpage
\clearpage
\begin{figure}
\plotone{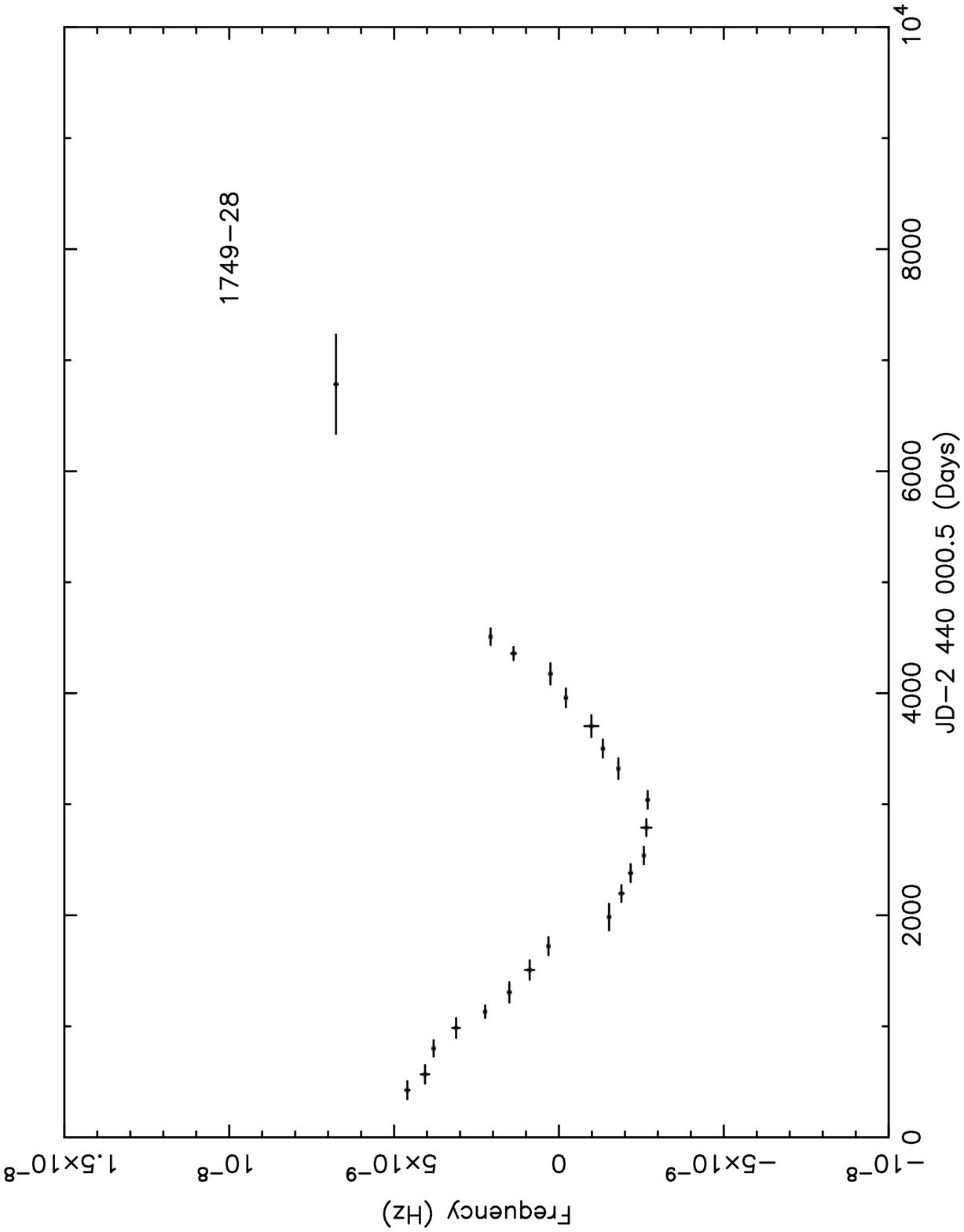}
\end{figure}
\newpage
\clearpage
\begin{figure}
\plotone{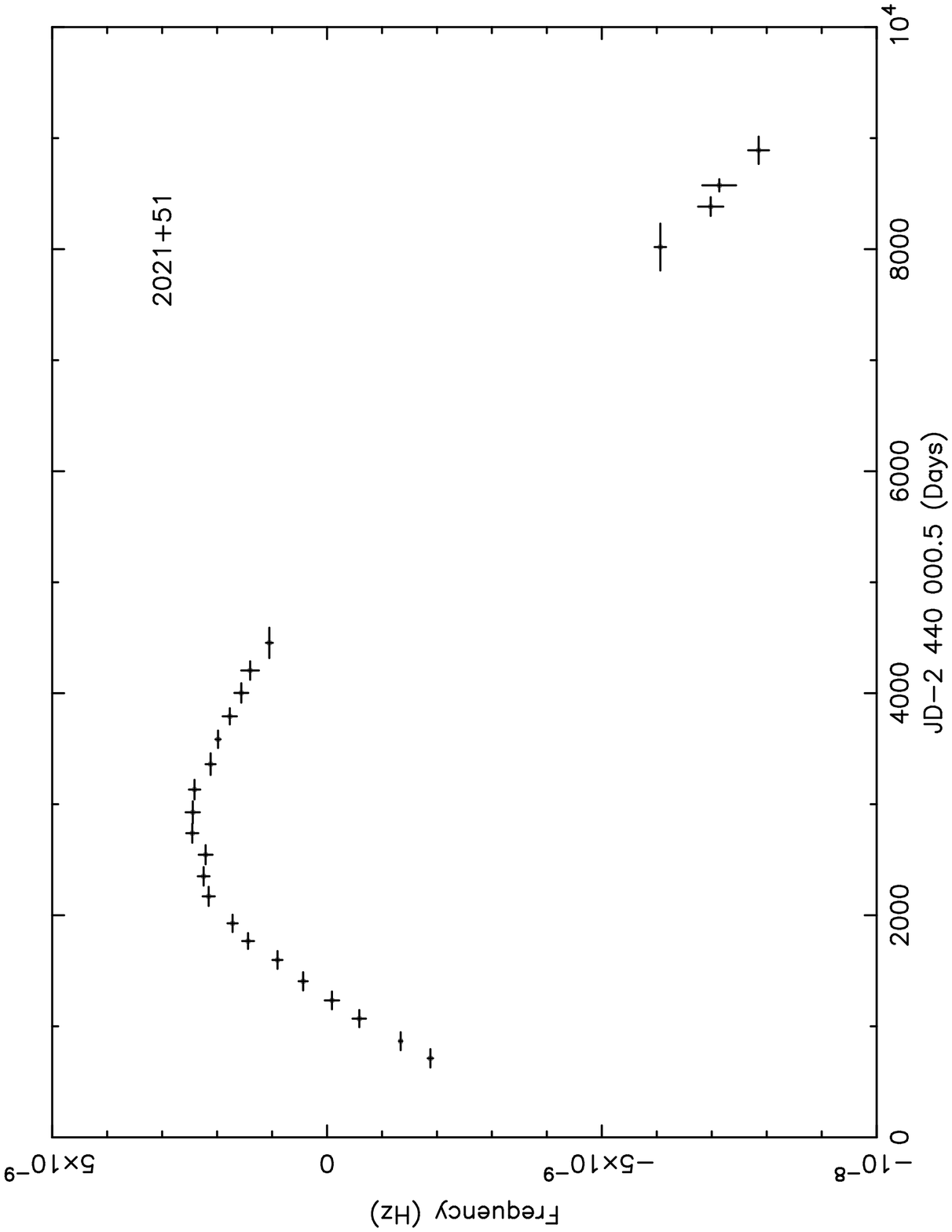}
\end{figure}
\begin{figure}
\plotone{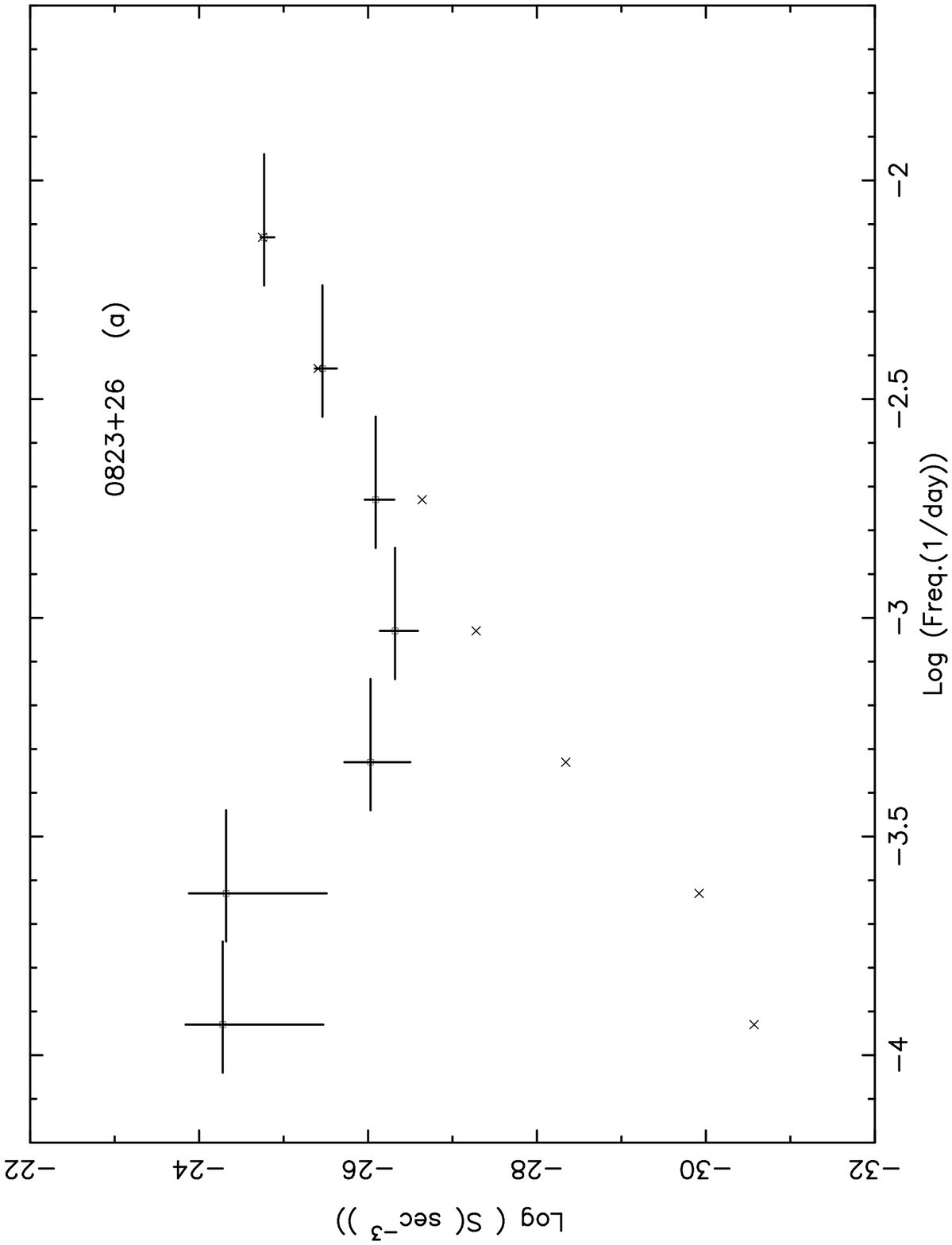}
\end{figure}
\begin{figure}
\plotone{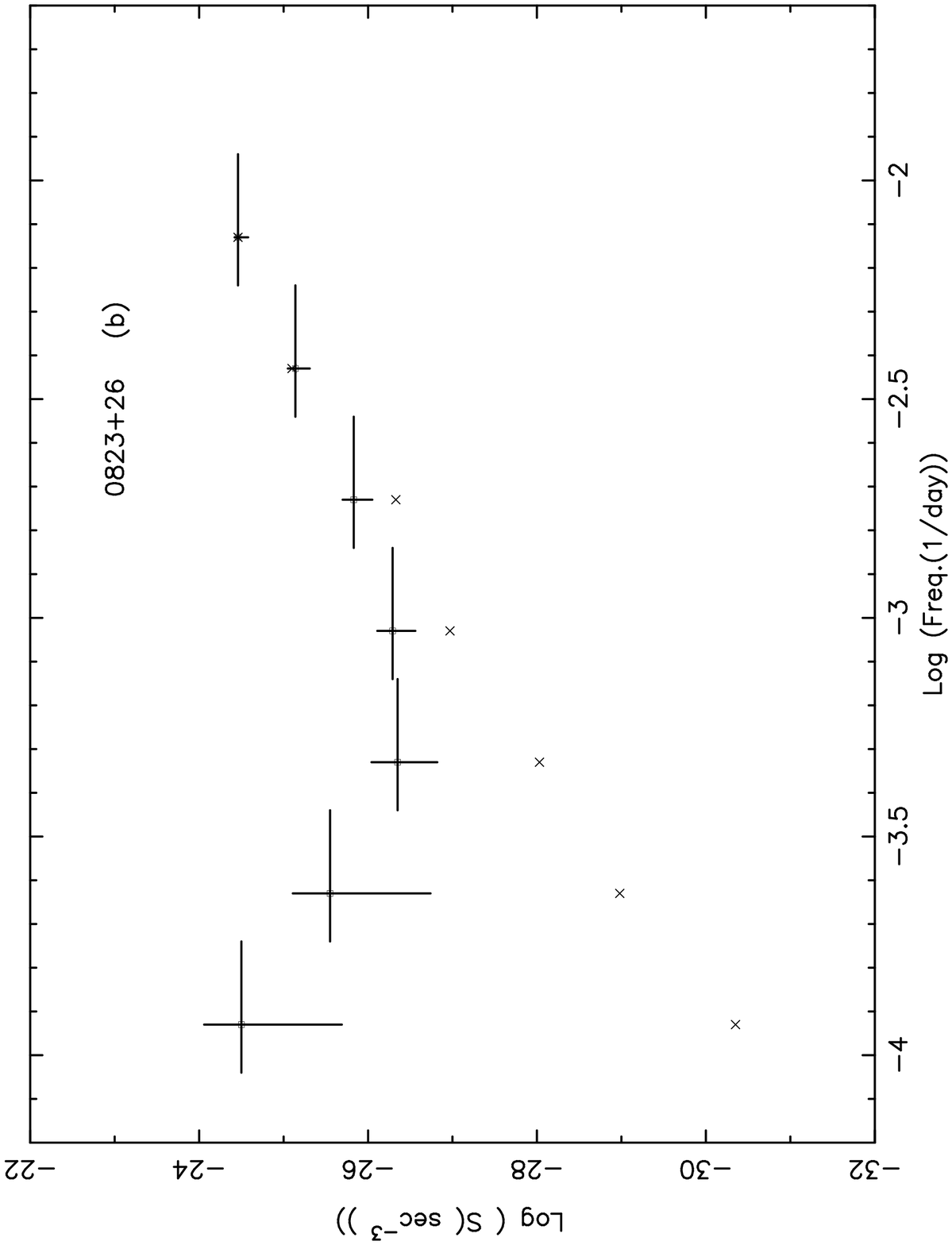}
\end{figure}
\begin{figure}
\plotone{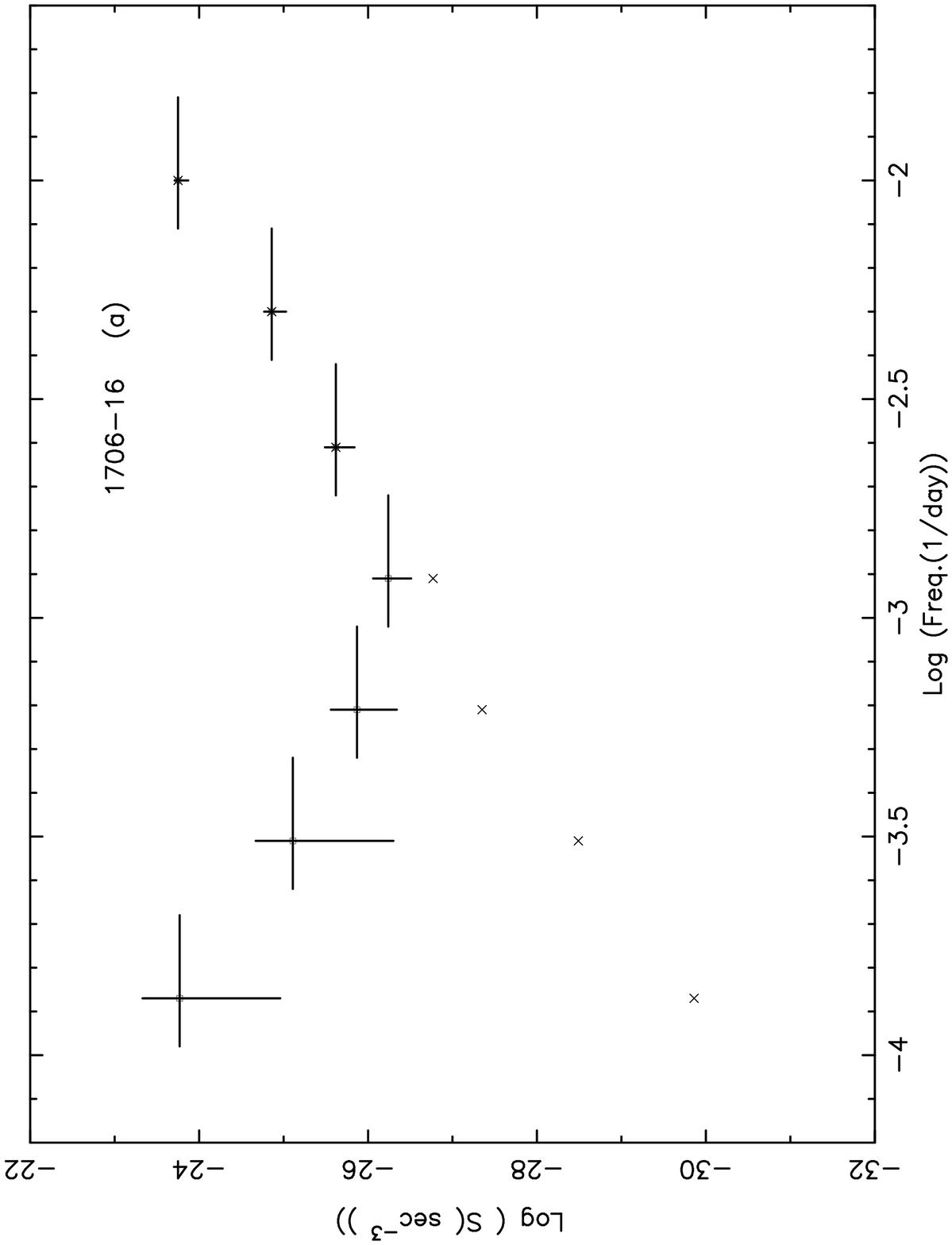}
\end{figure}
\begin{figure}
\plotone{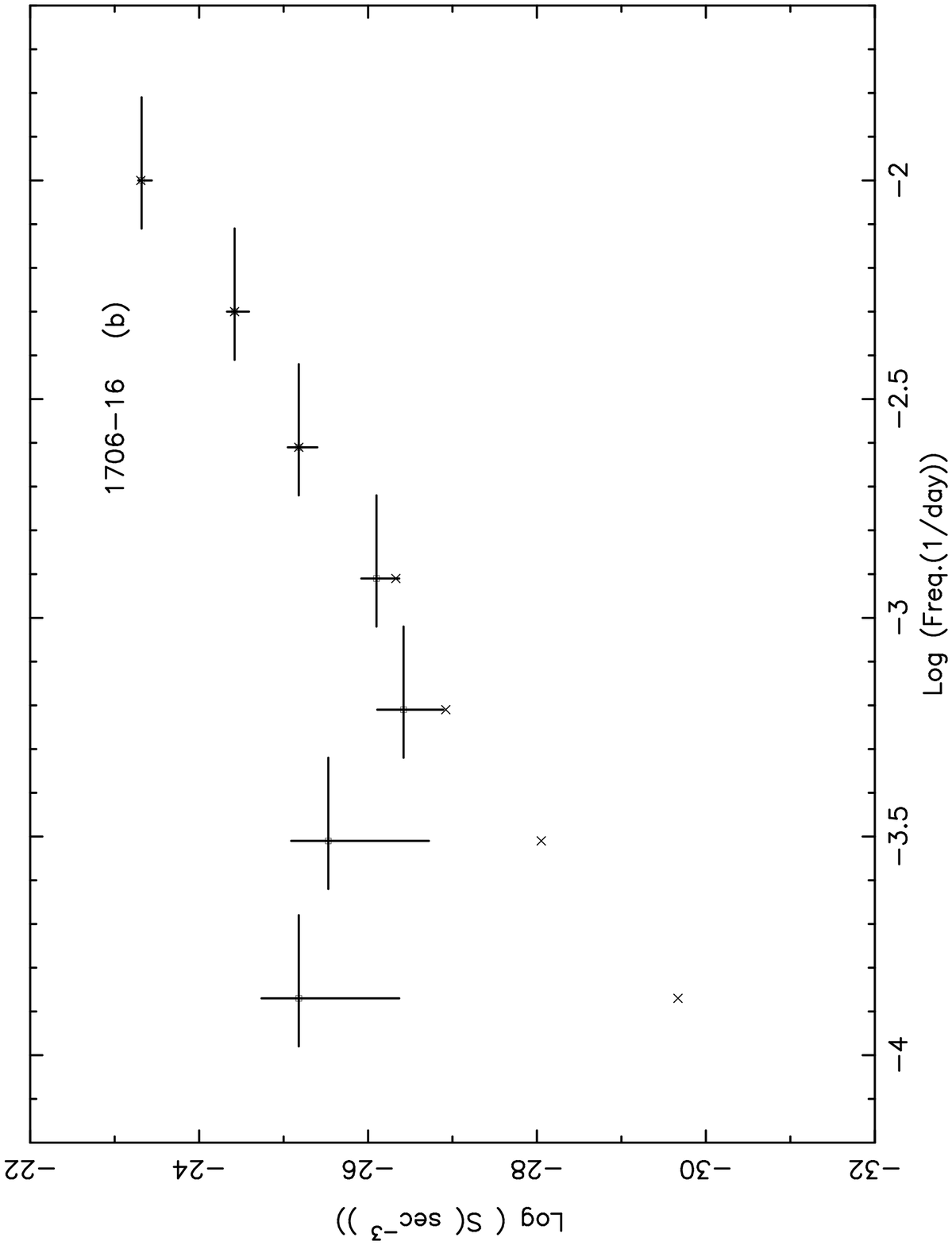}
\end{figure}
\begin{figure}
\plotone{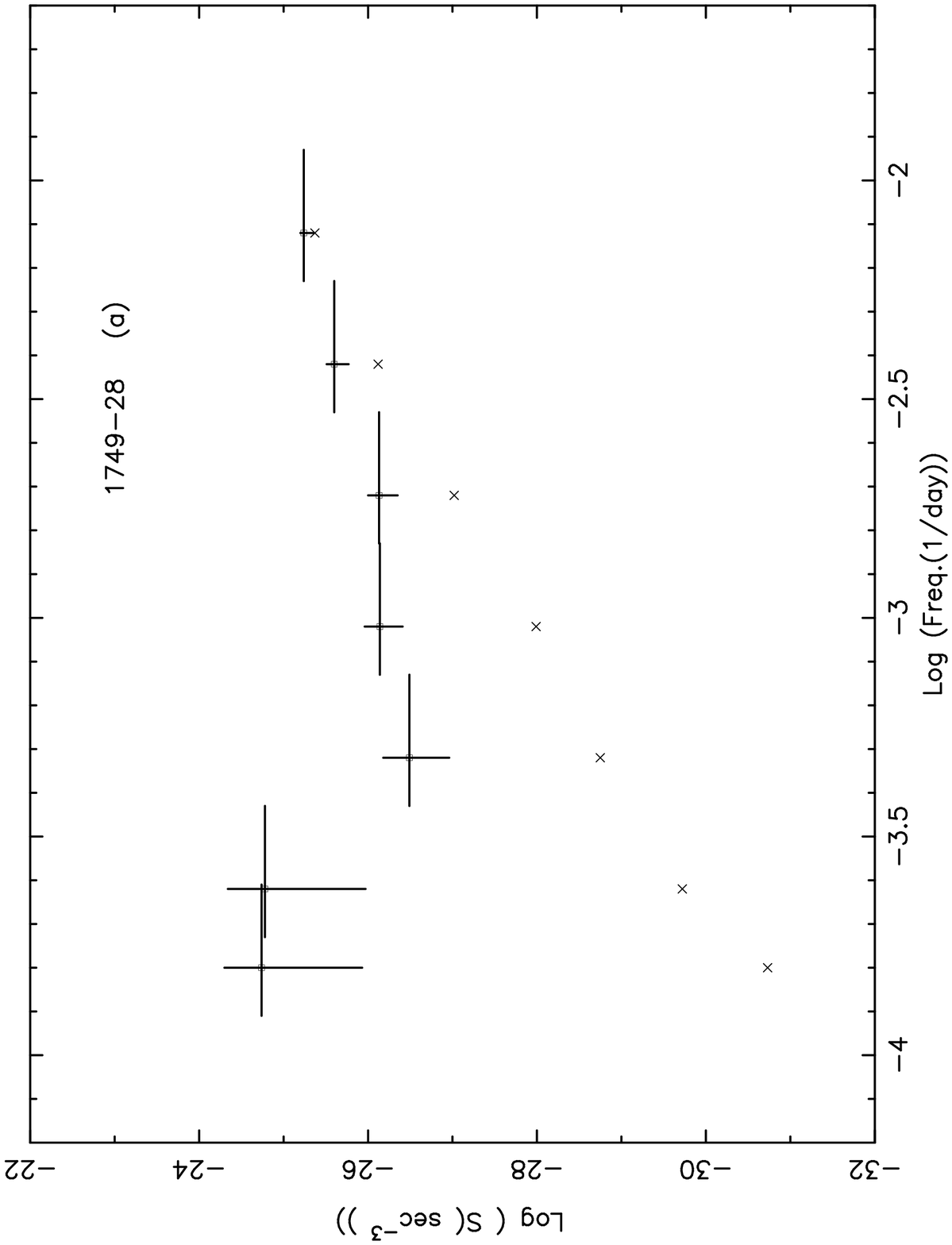}
\end{figure}
\begin{figure}
\plotone{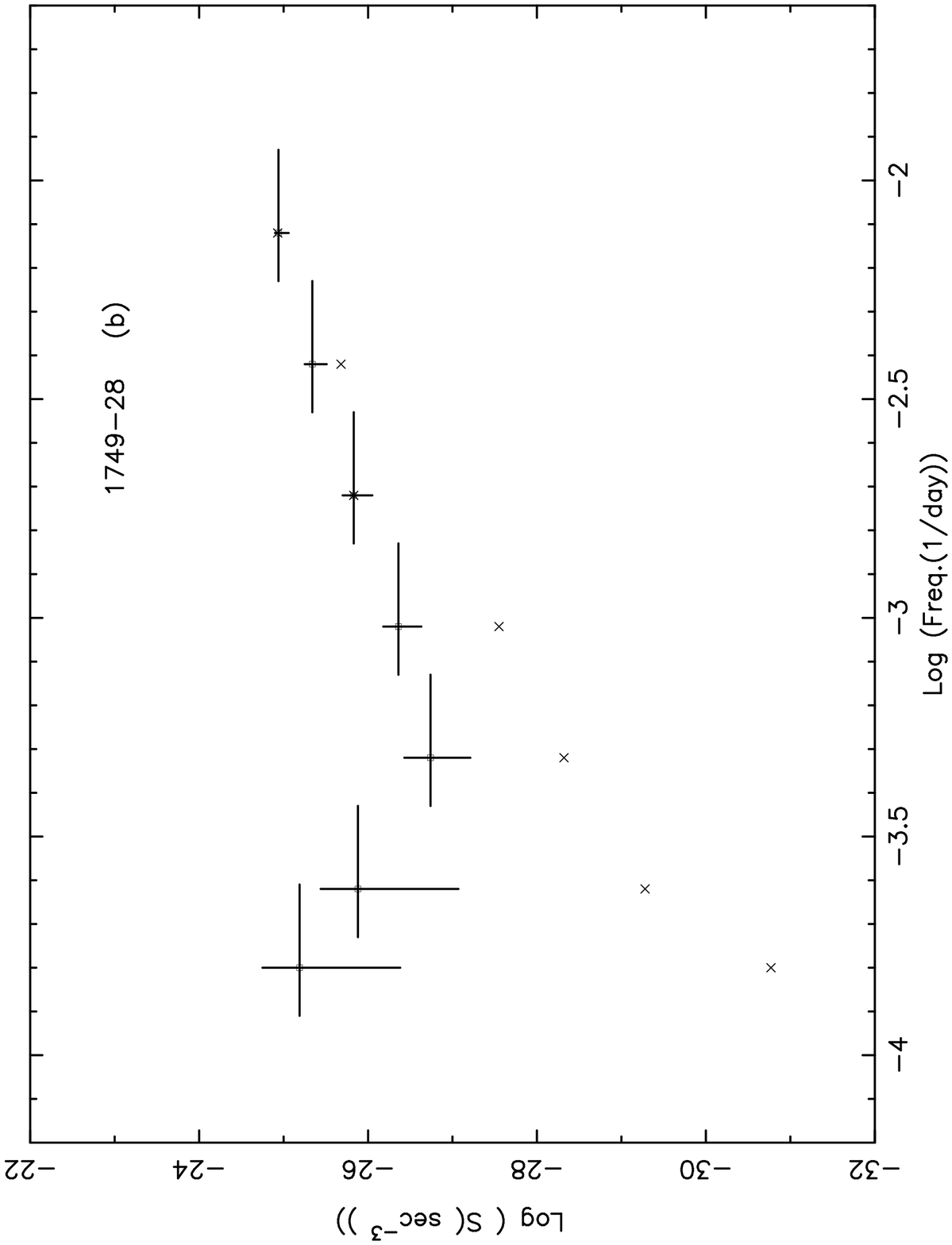} 
\end{figure}
\begin{figure}
\plotone{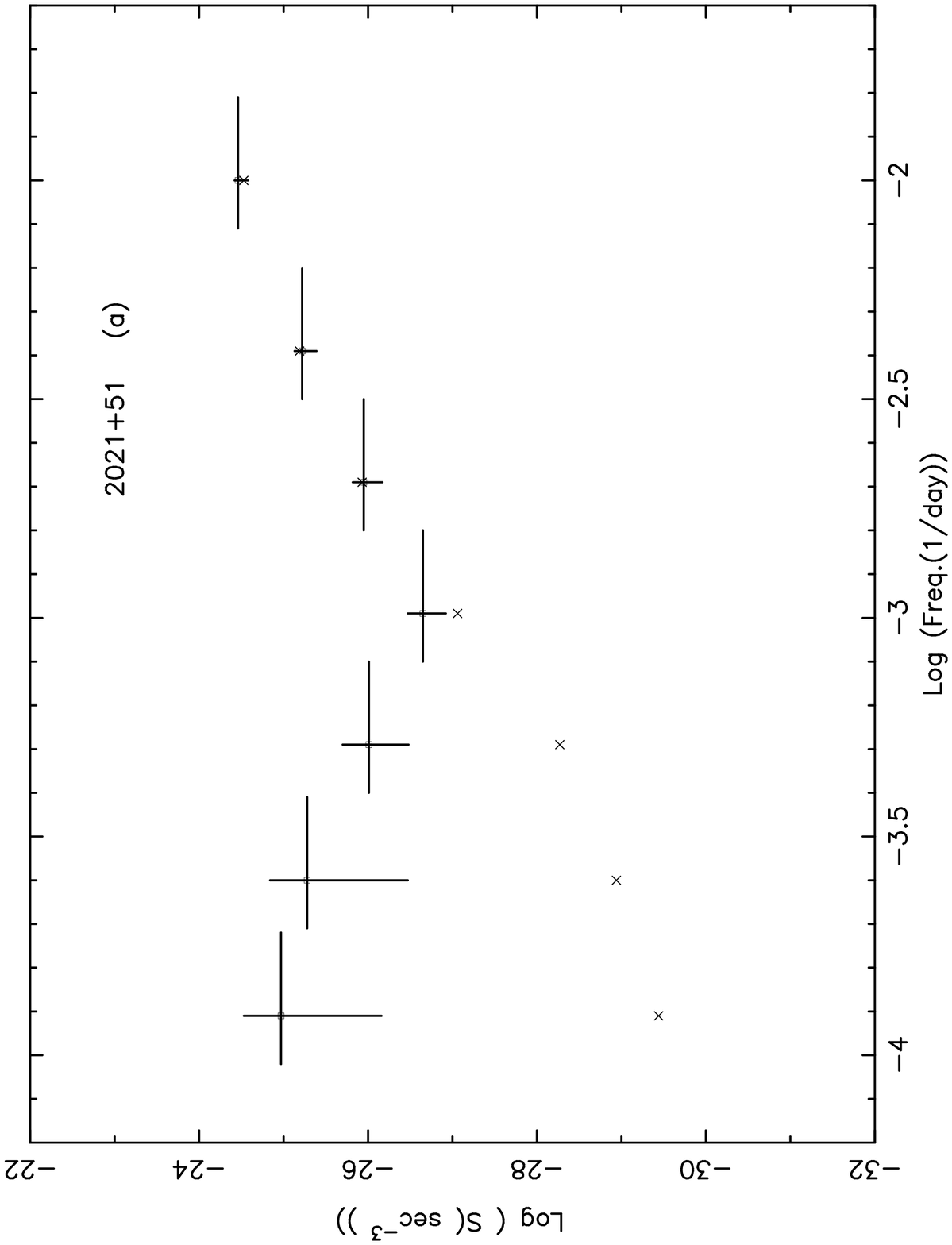}
\end{figure}
\begin{figure}
\plotone{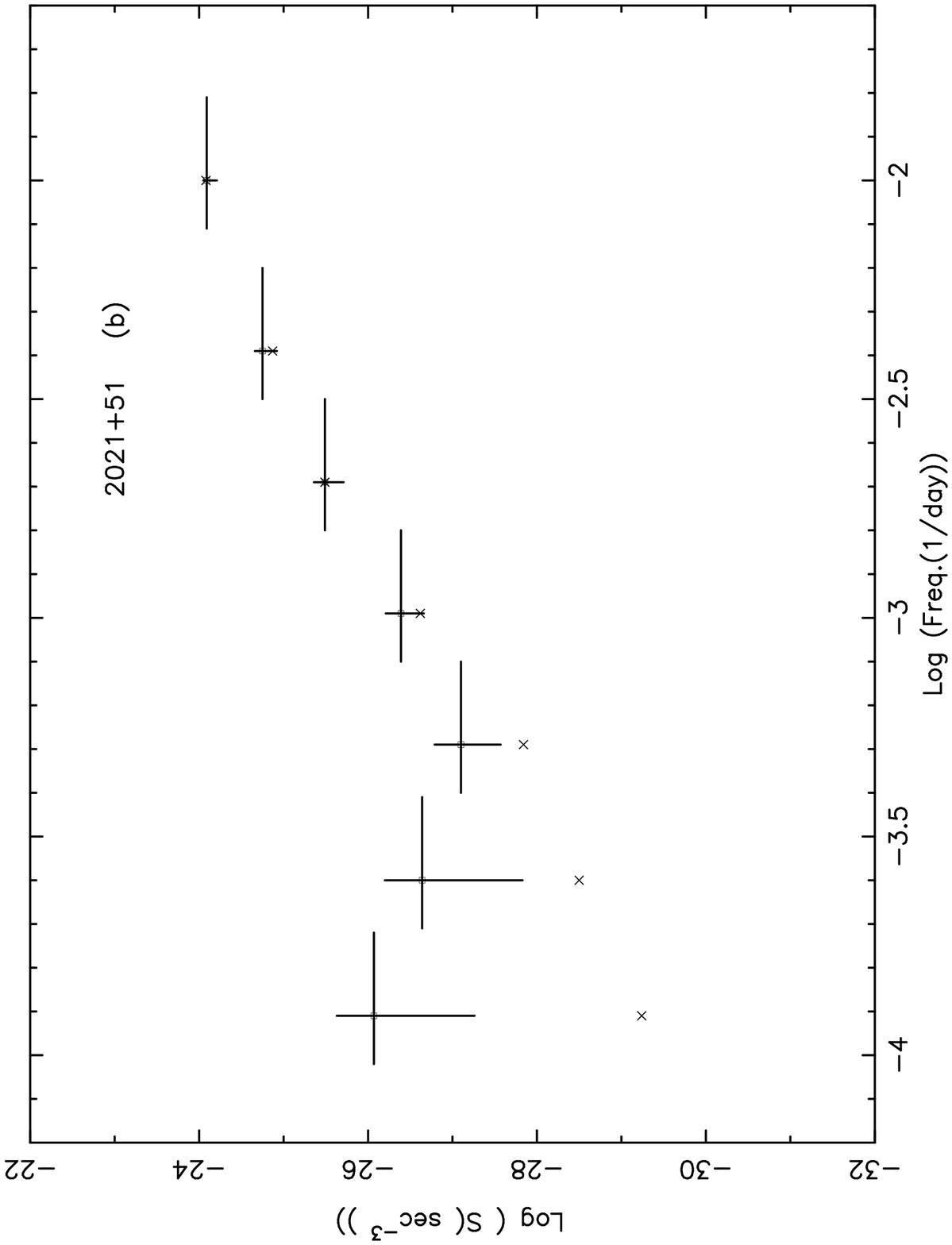}
\end{figure}

\end{document}